\documentclass[aps,pra,twocolumn,superscriptaddress]{revtex4-1}

\pdfoutput=1
\usepackage[utf8]{inputenc}
\usepackage[english]{babel}
\usepackage[T1]{fontenc}
\usepackage{amsmath}
\usepackage{hyperref}
\usepackage{tikz}
\usepackage{lipsum}
\usepackage{float}
\usepackage{graphicx}
\usepackage{latexsym}
\usepackage{colortbl}
\usepackage{multirow}
\usepackage{placeins}
\usepackage{longtable}

\usepackage{amssymb}
\usepackage{csquotes}
\usepackage{nicefrac}
\usepackage{amsfonts}
\usepackage{enumerate}
\usepackage{upgreek}
\usepackage{subfigure}
\usepackage{bm}
\usepackage[version=3]{mhchem}
\usepackage{nicefrac}
\usepackage{bbold}
\usepackage{verbatim}
\usepackage{multirow}
\usepackage{color}
\usepackage{comment}
\usepackage{xcolor}
\usepackage{soul}

\usepackage{braket}

\renewcommand{\vec}[1]{\mathbf{#1}}

\begin{document}

\title{Towards effective models for low-dimensional cuprates: \\
From ground state Hamiltonian reconstruction to spectral functions}

\author{Hannah Lange}
\affiliation{Ludwig-Maximilians-University Munich, Theresienstr. 37, Munich D-80333, Germany}
\affiliation{Max-Planck-Institute for Quantum Optics, Hans-Kopfermann-Str.1, Garching D-85748, Germany}
\affiliation{Munich Center for Quantum Science and Technology, Schellingstr. 4, Munich D-80799, Germany}

\author{Tizian Blatz}
\affiliation{Ludwig-Maximilians-University Munich, Theresienstr. 37, Munich D-80333, Germany}
\affiliation{Munich Center for Quantum Science and Technology, Schellingstr. 4, Munich D-80799, Germany}

\author{Ulrich Schollwöck}
\affiliation{Ludwig-Maximilians-University Munich, Theresienstr. 37, Munich D-80333, Germany}
\affiliation{Munich Center for Quantum Science and Technology, Schellingstr. 4, Munich D-80799, Germany}

\author{Sebastian Paeckel}
\affiliation{Ludwig-Maximilians-University Munich, Theresienstr. 37, Munich D-80333, Germany}
\affiliation{Munich Center for Quantum Science and Technology, Schellingstr. 4, Munich D-80799, Germany}

\author{Annabelle Bohrdt}
\affiliation{Ludwig-Maximilians-University Munich, Theresienstr. 37, Munich D-80333, Germany}
\affiliation{Munich Center for Quantum Science and Technology, Schellingstr. 4, Munich D-80799, Germany}

\date{\today}
\begin{abstract}
Understanding which minimal effective model captures the essential physics of cuprates is a key step towards unraveling the mechanism behind high-$T_c$ superconductivity. Recent measurements of the dynamical spin structure factor (DSF) in cuprate ladder compounds have indicated the presence of a large effective attraction in the single-band Hubbard model, possibly mediated by phonons. Here, we demonstrate that similar DSF features can also be captured by $t$-$J$ descriptions with or even without any attractive term. Motivated by this observation, we systematically investigate the strength and origin of different contributions to the single-band Hamiltonians by downfolding either from the three-band Emery model or the electron-phonon coupled Hubbard-Holstein model. For one-dimensional systems, we find that the extended versions of both single-band descriptions can reproduce the experimentally observed DSF signatures. Finally, we extend our analysis to two dimensions by comparing two-hole correlation functions for the different single-band models. Our results provide new insights into the long-standing question of which single-band Hamiltonian can capture the essential physics of cuprates.
\end{abstract}
\maketitle

The discovery of superconductivity in LaBaCuO~\cite{Bednorz1986} has inspired decades of extensive experimental and theoretical research into cuprate materials. The common belief is that the relevant physics takes place in the copper oxide planes, with each
CuO$_2$ unit cell consisting of a copper $d_{x^2-y^2}$ orbital and two oxygen $p_{x,y}$ orbitals, which is described by the three-band Emery model~\cite{Emery1987}. However, this model is very challenging to simulate for most methods -- although methods like dynamical mean-field theory~\cite{Weber2010,Weber_2012} allow for multi-band simulations under the assumption of local interactions and correlations~\cite{Georges1996} -- and, hence, are often downfolded to a single-band description. However, which single-band model can most accurately capture the cuprate physics, how large the corresponding Hamiltonian parameters are, which terms are required and which terms can be neglected, has remained unclear until today. After decades of study, this question has recently regained attention, as large-scale numerical simulations suggest that additional Hamiltonian terms might be essential for capturing the cuprate physics~\cite{Jiang2021,Xu2024}. Furthermore, increasing experimental evidence also points toward the need for more complex models beyond the plain-vanilla Hubbard model \cite{Padma2025,Scheie2025,Li2025}. Thus, novel downfolding procedures yielding effective lattice models with potentially long-ranged interactions can provide new insights into the relevant physics of real materials~\cite{Jiang2021}.\\

In the earliest works, downfolding was typically done by constructing a new basis set of wave functions that are centered at the copper sites. In a perturbative analysis, Zhang and Rice argued that upon doping with a single hole, this new basis consists of additional holes delocalized over the $p$-sites, forming singlets with the neighboring $d$-hole~\cite{Zhang1988}. These so-called \textit{Zhang-Rice singlets} have also been observed, both experimentally and numerically, see e.g. Refs.~\cite{Ye2013,ye2023visualizingzhangricesingletmolecular,Tjeng1997,Harada2002,Li2021}. In the limit of small hopping integrals between the orbitals, one arrives at the single-band $t$-$J$ model,
\begin{align}   
\hat{\mathcal{H}}_{tJ} =&\hat{\mathcal{P}}_G\left[\hat{\mathcal{H}}_t+\hat{\mathcal{H}}_{3s}\right]\hat{\mathcal{P}}_G +  J \sum_{\langle i,j \rangle} (\hat{\mathbf{S}}_i \cdot \hat{\mathbf{S}}_j - \frac{1}{4} \hat{n}_i \hat{n}_j) ,
\label{eq:tJ}
\end{align}
with the Gutzwiller projector $\hat{\mathcal{P}}_G$ projecting on singly occupied sites and the kinetic terms $\hat{\mathcal{H}}_t=-t\sum_{\langle i,j \rangle, \sigma} \left( \hat{c}^{\dagger}_{i\sigma} \hat{c}_{j\sigma} + \text{h.c.} \right) $ and $\hat{\mathcal{H}}_{3s}=- t_3 \sum_{\langle i,j,k \rangle, \sigma} \left( \hat{c}^\dagger_{k\sigma} \hat n_{j\bar{\sigma}} \hat{c}_{i\sigma} - \hat{c}^\dagger_{k\sigma} \hat{c}^\dagger_{j\bar{\sigma}} \hat{c}_{j\sigma} \hat{c}_{i\bar{\sigma}} + \text{h.c.} \right)$. We denote indices corresponding to neighboring lattice sites by $\langle \cdot,\cdot\rangle$. Similar results have been obtained 
e.g. in Refs.~\cite{Schuettler1992,Jefferson1992,Belinicher1993,Belinicher1994,Feiner1996,Feiguin2023}. In some of the works, additional terms like density-density interactions, further-range or density-assisted hopping terms are introduced~\cite{Xu2024,Jiang2021}. \\

When allowing for larger occupations, an effective single-band Fermi-Hubbard model can be derived~\cite{Schuettler1992,Feiner1996,Feiguin2023,Jiang2021},
\begin{align}  
\hat{\mathcal{H}}_\mathrm{FH}= -t \sum_{\langle i,j \rangle, \sigma} \left(\hat{c}_{i\sigma}^\dagger \hat{c}_{j\sigma} + \text{h.c.}\right) 
+ U \sum_{i} \hat{n}_{i\uparrow} \hat{n}_{i\downarrow}
\label{eq:FH}.
\end{align}

\begin{figure*}
\centering
\includegraphics[width=0.99\textwidth]{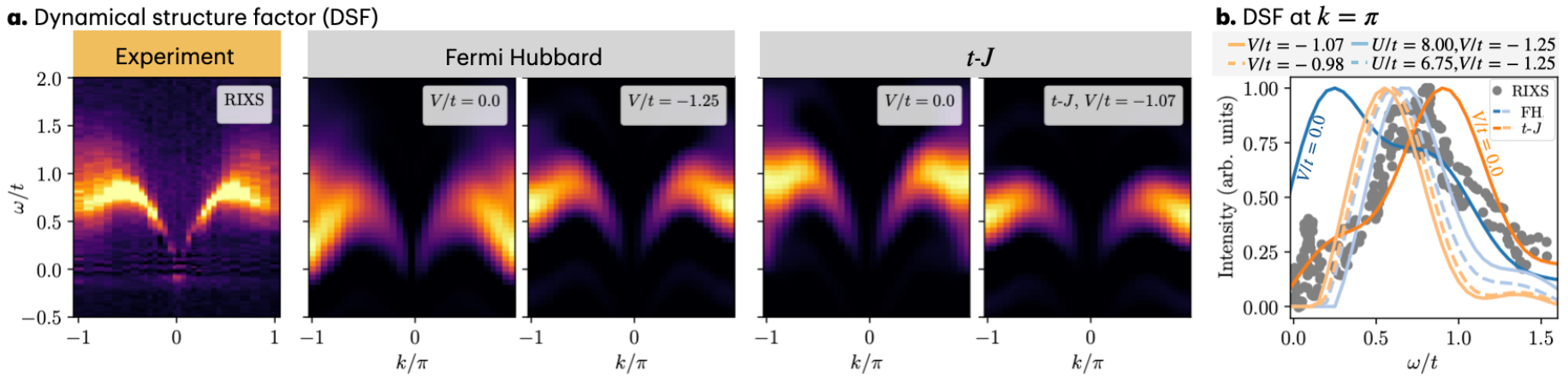}
\caption{The effective model for cuprate ladders. \textbf{a.} We compare the dynamical structure factor (DSF) obtained from RIXS measurements on Sr$_{14}$Cu$_{24}$O$_{41}$~\cite{Padma2025} (left) to four different effective model candidates: The Fermi-Hubbard model without and with attractive interactions $V/t=-1.25$ and rescaled $U/t=8+V$ (FH and FH$+V$), the bare $t$-$J$ model without attraction and the $t$-$J$ model with attraction $V/t=-1.07$ ($t$-$J+V$) for ladders of length $L=33$ and $\delta = 6\%$ hole doping (left to right). \textbf{b} FH$+V$ (light blue), bare $t$-$J$ (dark orange) and $t$-$J+V$ (light orange) models agree well with recent RIXS measurements~\cite{Padma2025}, as shown for exemplary momentum $k=\pi$. Dashed lines refer to FH$+V$ and reconstructed $t$-$J+V$ without rescaled on-site interaction $U/t=8$.  
}
\label{fig:LadderDSF}
\end{figure*}


Intriguingly, recent comparisons of resonant inelastic X-ray scattering (RIXS) and angle-resolved photoemission spectroscopy (ARPES) on one-dimensional (1D) cuprates with numerics for the single-band Hubbard model have shown that attractive interactions ($V_d<0$) for neighbors with distance $d$ of the form
\begin{align}
    \mathcal{\hat{H}}_V=\sum_{\langle i,j\rangle_d}V_d\,\hat{n}_i\hat{n}_j,
    \label{eq:V}
\end{align}
need to be included to account for the observed features of the dynamical structure factor (DSF) and the photoemission spectrum, mostly limited to nearest-neighbor $V:=V_{d=1}$ terms on the order of $V\approx-t$. In this work, also $V^\prime:=V_{d=2}$ will be considered. This attraction could potentially be induced by electron-phonon interactions~\cite{Li2025,Tohyama2024,Tang2024,Shen2024,Chen2021,Wang2021,Feiguin2023}. For ladder compounds, RIXS and neutron scattering measurements have shown the necessity to include similar attractive interactions very recently~\cite{Scheie2025,Padma2025}.  \\

In this work, we systematically analyze the Fermi-Hubbard model and compare it to the $t$–$J$ model, both with and without additional terms beyond the plain-vanilla versions like  attraction or density-assisted hopping. We begin by investigating 1D and ladder compounds, focusing on the origin of the potential attractive interactions and their impact on the spectral properties. Then, we extend our analysis to the 2D case. Specifically, we $(i)$ compare the DSF for ladder compounds within the $t$–$J$ and Fermi-Hubbard models with and without nearest-neighbor attraction, finding that both reproduce spectral features consistent with resonant RIXS measurements; $(ii)$ explore the microscopic origin of the attractive interaction as well as other additional terms by applying ground-state reconstruction schemes based on the Hubbard–Holstein model with electron-phonon coupling and three-band Fermi-Hubbard models; $(iii)$ validate the reconstructed models by computing the DSF, which again shows good agreement for both the FH$+V$ and $t$–$J$ models; and $(iv)$ extend the analysis to two dimensions by comparing correlation functions for the different effective one-band models. \\

\textit{Ladder compounds.--} As a general motivation for our work, we consider the recent experiments probing the dynamical spin structure factor (DSF) of the cuprate ladder compounds on Sr$_{14}$Cu$_{24}$O$_{41}$~\cite{Padma2025} and Sr$_{2.5}$Ca$_{11.5}$Cu$_{24}$O$_{41}$~\cite{Scheie2025} at hole doping $\delta=6\%$. The magnetic excitations of these systems are theoretically well understood: in an isotropic, undoped ladder, the ground state consists of spin singlets on the rungs, and the elementary magnetic excitations are triplets, leading to a continuum of two-triplon excitations~\cite{Barnes1993}. Upon doping, holes break spin singlets and give rise to quasiparticles with loosely bound spin and charge components, which in turn contribute spin-flip excitations to the DSF. In Fig.~\ref{fig:LadderDSF}\textbf{a}, we show the RIXS measurement from Ref.~\cite{Padma2025} at $260\,\mathrm{K}$ (left panel) at hole doping $\delta=6\%$, where the two-triplon continuum is clearly visible, while spin-flip excitations are strongly suppressed. This behavior is in stark contrast to results from a bare Fermi-Hubbard ladder of length $L_x=33$ (see Fig.~\ref{fig:LadderDSF}\textbf{a}, second panel), obtained via density matrix renormalization group (DMRG) using hopping parameters $t_\perp/t=0.84$ and  $t_\mathrm{diag}/t=-0.3$ as extracted in Ref.~\cite{Padma2025}. In the Fermi-Hubbard case, prominent spin-flip excitations with downward dispersion near the Brillouin zone boundary appear below the two-triplon continuum -- features that are absent in the experiment. Notably, the three-band Hubbard model has also been shown to not capture this suppression~\cite{Li2021}. To reconcile theory and experiment, Refs.~\cite{Padma2025, Scheie2025} proposed the inclusion of a nearest-neighbor attraction of strength  $V/t=-1.25$ along both ladder directions and a rescaled on-site interaction $U/t=8+V$. The attraction $V$ increases the rung-pairing correlations  significantly~\cite{Padma2025, Zhou2023}, and the stronger pairing suppresses the spin-flip contributions (Fig.~\ref{fig:LadderDSF}\textbf{a}, 3rd panel). However, the bare $t$-$J$ model -- without any additional attraction -- naturally features strong pairing for holes in different legs of the ladder~\cite{Tsunetsugu1994,White1997}. As we show here, this intrinsic pairing of the bare $t$-$J$ ladder gives rise to a similar pairing strength as for the FH+$V$ ladder, and consequently to very similar features in the DSF: spin-flip excitations are suppressed, and the energy of the spectral peak at the Brillouin zone boundary $k=\pi$ aligns closely with that observed in experiment (see Fig.~\ref{fig:LadderDSF}\textbf{a} 4th panel and Fig.~\ref{fig:LadderDSF}\textbf{b}). 

In general, the spectral weight is shifted towards $\pi/2$ for the RIXS, which is not the case for the inelastic neutron scattering results in Ref.~\cite{Scheie2025}. More information on the theoretical DSF can be found in the Supplemental Material (SM)~\cite{SM}.

The similar pairing strength for both FH$+V$ and $t$-$J$ models can be quantified in terms of the binding energies $E_B=2E(N^h-1)-E(N^h)-E(N^h-2)$ (where $N^h=\delta L_xL_y$ is the number of holes): For $L_x=64$, the FH$+V$ ladder (with $t_\mathrm{diag}=0$) features binding energies $E_B^{\mathrm{FH}+V}/t=0.23$, which is comparable to those of the bare $t$-$J$ ladder (with $J_\parallel/t=0.5$, $J_\perp/t=0.35$, $J_\mathrm{diag}=t_\mathrm{diag}=0$), $E_B^{tJ}/t=0.17$. Both are much larger than $E_B^{\mathrm{FH}}/t=0.09$ for the bare Fermi-Hubbard ladder. 

To enable a more systematic comparison between the different single-band models, we employ a reconstruction scheme to derive an effective $t$-$J$ Hamiltonian that reproduces the ground state of the FH$+V$ ladder with and without rescaled $U$, see SM. We find that the interaction term $V$ is suppressed to $V/t=-1.07$ and $V/t=-0.98$, respectively. In both cases, the resulting dynamical structure factor (DSF) closely resembles the DSF for the FH$+V$ model, see see Fig.~\ref{fig:LadderDSF}\textbf{a}(right). 

These observations show that both the Fermi-Hubbard model with attraction (FH$+V$) and the considered $t$-$J$ models with and without attraction appear to be suitable effective single-band models for the cuprate ladder compounds. Due to their close connection to two-dimensional compounds, this raises a more fundamental question: What range, strength, and sign of interactions should be included in an effective model to faithfully capture the essential physics of cuprate materials?\\

\textit{One-dimensional compounds.--} In the following, we address this question by employing ground state reconstruction schemes to determine the best effective single-band description from the three-band and Hubbard-Holstein models, and comparing their spectral features to RIXS results from the 1D cuprate compound Ba$_2$CuO$_{2+\delta}$~\cite{Li2025}. 
More precisely, for the three-band model, we consider the hole-doped three-band model 
\begin{align}   
\hat{\mathcal{H}}_{3b}=& -t_{pd} \sum_{\langle ij \rangle \sigma} (\hat{d}^{\dagger}_{i\sigma} \hat{p}_{j\sigma} + \text{h.c.}) 
- t_{pp} \sum_{\langle\langle ij \rangle\rangle \sigma} (\hat{p}^{\dagger}_{i\sigma} \hat{p}_{j\sigma} + \text{h.c.}) \notag \\
&+\sum_{\mu\in\{p,d\}} U_\mu \sum_i \hat{n}^\mu_{i\uparrow} \hat{n}^\mu_{i\downarrow} 
+ \Delta_{pd} \sum_{i\sigma} \hat{p}^{\dagger}_{i\sigma} \hat{p}_{i\sigma},
\label{eq:3band}
\end{align}
with orbital-dependent on-site interactions $U_{d,p}>0$ and hoppings $t_{pd,dd}$, as well as a potential offset of the oxygen orbitals $0<\Delta_{pd}<U_d$. The annihilation (creation) operators $\hat{d}^{(\dagger)}$ and $\hat{p}^{(\dagger)}$ create holes on copper and oxygen sites respectively. The half-filled case ($n=1$) is defined by one hole per unit cell.

\begin{figure}[t]
\centering
\includegraphics[width=0.44\textwidth]{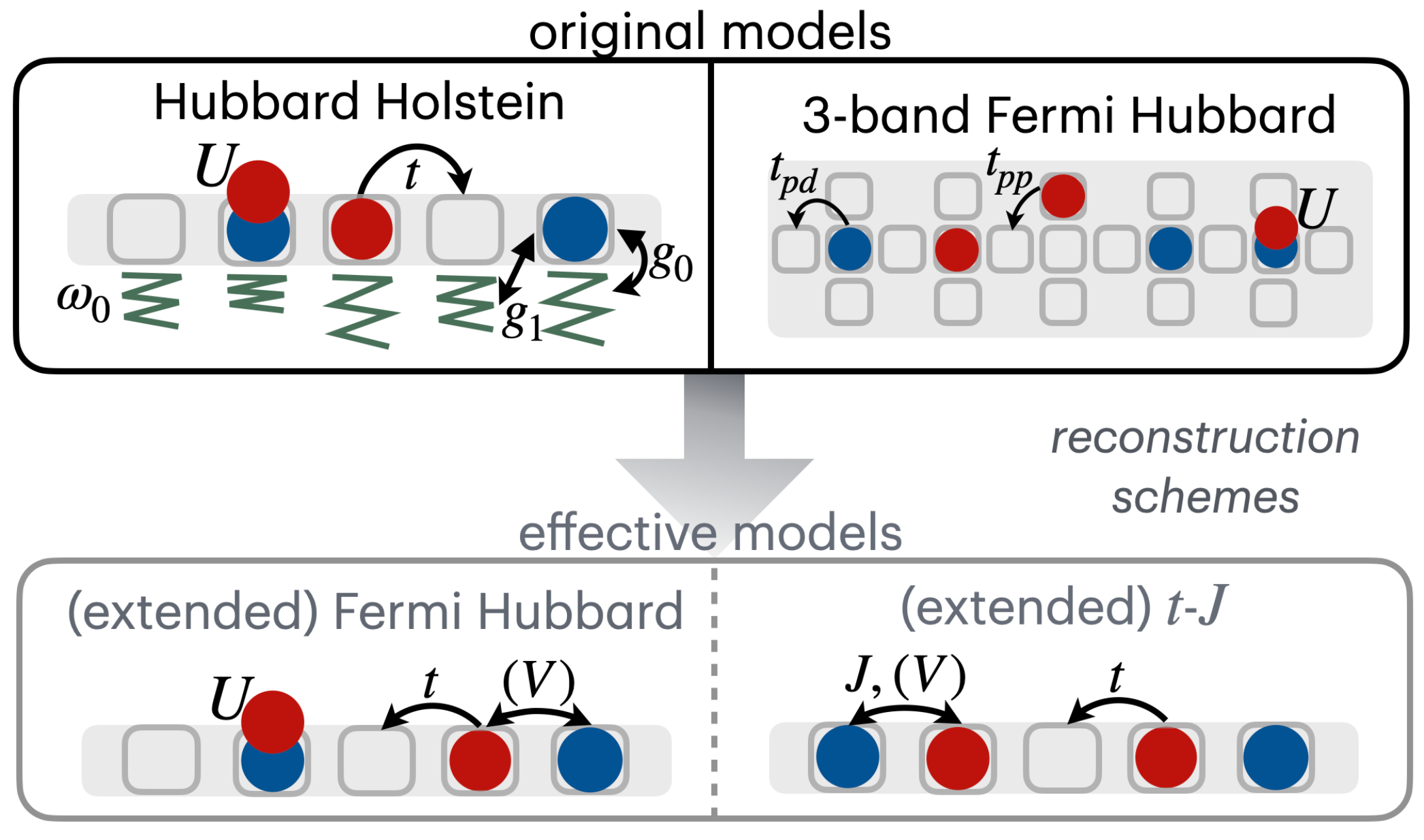}
\caption{ Ground state reconstruction schemes: We start from Hubbard-Holstein or three-band Fermi-Hubbard chains (top) and reconstruct the corresponding effective single-band Fermi-Hubbard or $t$-$J$ models (bottom). }
\label{fig:ReconstructionScheme}
\end{figure}

\begin{table}[t]
\centering
\begin{tabular}{|c||c|c|c|c||c|c|c|}
\hline
& \multicolumn{4}{c||}{to FH} &  \multicolumn{3}{c|}{to $t$-$J$} \\
\multirow{-2}{*}{Reconstr. from} & $U/t$ & $t_n/t$ & $V/t$ &$V^\prime/t$ & $J/t$ & $t_3/t$ & $V/t$ \\ \hline\hline

\multirow{2}{*}{HH} 
& 8.05 & - & - & - & 0.51 & 0.06 & - \\ 
& 7.64 & - & -0.64 & -0.34 & 0.52 & 0.05 & -0.44 \\ \hline

\multirow{2}{*}{3-band FH} 
& 11.32 & 0.64 & -  & - & 0.68 & 0.08 & - \\ 
& 11.41 & 0.65 & 0.07 & - & 0.66 & 0.08 & -0.02 \\ \hline

\end{tabular}
\caption{Reconstructed parameters for ground state Hamiltonian reconstruction of Hubbard-Holstein (HH) and three-band Fermi-Hubbard (FH) chains to effective single-band Fermi-Hubbard and $t$-$J$ models without and with a nearest-neighbor density interaction Eq. \eqref{eq:V} (top vs. bottom). We omit the next-nearest neighbor interaction $V^\prime$ for the $t$-$J$ reconstruction here , since tit is one order of magnitude smaller than for FH, see SM~\cite{SM}. for All parameters are extrapolated to infinite system sizes, see SM~\cite{SM}.}
\label{tab:Params}
\end{table}

For the Hubbard–Holstein model, we include electron–phonon interactions given by the extended Hubbard-Holstein Hamiltonian
\begin{align}   \hat{\mathcal{H}}_\mathrm{HH}=&\hat{\mathcal{H}}_\mathrm{FH}
+ \omega_0 \sum_i \hat{b}_i^{\dagger} \hat{b}_i
+ g_0 \sum_i \hat{n}_i\left( \hat{b}_i^{\dagger} + \hat{b}_i \right) \notag \\
&+ g_1 \sum_{\langle i,j\rangle}\hat{n}_i\left( \hat{b}_j^{\dagger} + \hat{b}_j \right) 
\label{eq:HH},
\end{align}
with phonon operators $\hat{b}^{(\dagger)}$, phonon frequency $\omega_0$, and on-site (nearest-neighbor) electron-phonon coupling $g_{0(1)}$. The strength of the resulting effective attraction, and its impact on spectral features has been considered e.g. in Refs.~\cite{Wang2021,Li2025,Shen2024,Tohyama2024}, suggesting an effective attraction on the order of $-t$, similar to that found in ladder compounds. Note that despite the effective attraction that can be derived from this model, it can not explain the emergence of unconventional superconductivity due to the large effective mass of the bipolaronic 2-electron bound states~\cite{Chakraverty1998}.  \\

The reconstruction we perform proceeds as follows: Starting from ground states of $(i)$ the Hubbard-Holstein model~\eqref{eq:HH} with $\omega_0/t=0.2$, $g_0/t=0.3$, $g_1/t=0.15$ (as estimated in Ref.~\cite{Wang2021}) and $(ii)$ the three-band Emery model \eqref{eq:3band} with $t_{pp}/t_{pd}=0.5$, $U_d/t_{pd}=6.0$, $U_p/t_{pd}=3.0$, $\Delta_{pd}/t_{pd}=3.5$ as used in Ref.~\cite{Jiang2021}, we construct the effective FH$+V$ and $t$-$J$($+V$) Hamiltonians, see Fig.~\ref{fig:ReconstructionScheme}. For the reconstruction from $(i)$, this amounts to finding the effective model for which the energy variance measured w.r.t. the original ground state is minimal~\cite{Qi2019}. This condition can be rewritten in terms of a correlation matrix, with the eigenvalues corresponding to the variance of an effective Hamiltonian given by the corresponding eigenvector. In analogy to a Lang-Firsov transformation of the Hubbard-Holstein Hamiltonian, where additionally a next-nearest neighbor interaction $V^\prime$ arises, we also include this term in this case. For $(ii)$, we construct the Wannier functions centered at the copper site in order to construct a unitary transformation from the three-band to the single-band description~\cite{Jiang2021} and apply the reconstruction scheme from $(i)$ afterwards. Further details can be found in the SM \cite{SM}.\\

The reconstructed parameters are presented in Tab.~\ref{tab:Params}. For both original models, we find a strong on-site interaction $U\gg t$ in the single-band Fermi-Hubbard model, corresponding to a spin exchange of approximately $J/t\approx 0.5$. Additionally, the phononic description based on the Hubbard-Holstein model indeed yields an effective attraction in both the Fermi-Hubbard and $t$-$J$ models, though its amplitude is smaller than anticipated for the ladder systems discussed above. In contrast, the protocol starting from the three-band model results in only a very weak density-density interaction, which is even repulsive in the Fermi-Hubbard model. Instead, consistent with Ref.~\cite{Jiang2021}, we observe a strong density-assisted hopping term in the Fermi-Hubbard model,
\begin{align}
    \mathcal{\hat{H}}^\mathrm{kin}_n=-t_n\sum_{\langle i,j\rangle\sigma} \left(\hat{c}_{i\sigma}^\dagger\hat{c}_{j\sigma}+\mathrm{h.c.} \right)\hat{n}_{i\bar{\sigma}}.
\end{align}
The corresponding on-site repulsion looks stronger than usual at first sight, but when normalized to all hopping contributions $U/(t_n\langle n\rangle +t)\approx 8$, as usually considered. This renormalized hopping strength also explains the stronger $J$ in the lower right part of Tab.~\ref{tab:Params}.\\

\begin{figure}[t]
\centering
\includegraphics[width=0.49\textwidth]{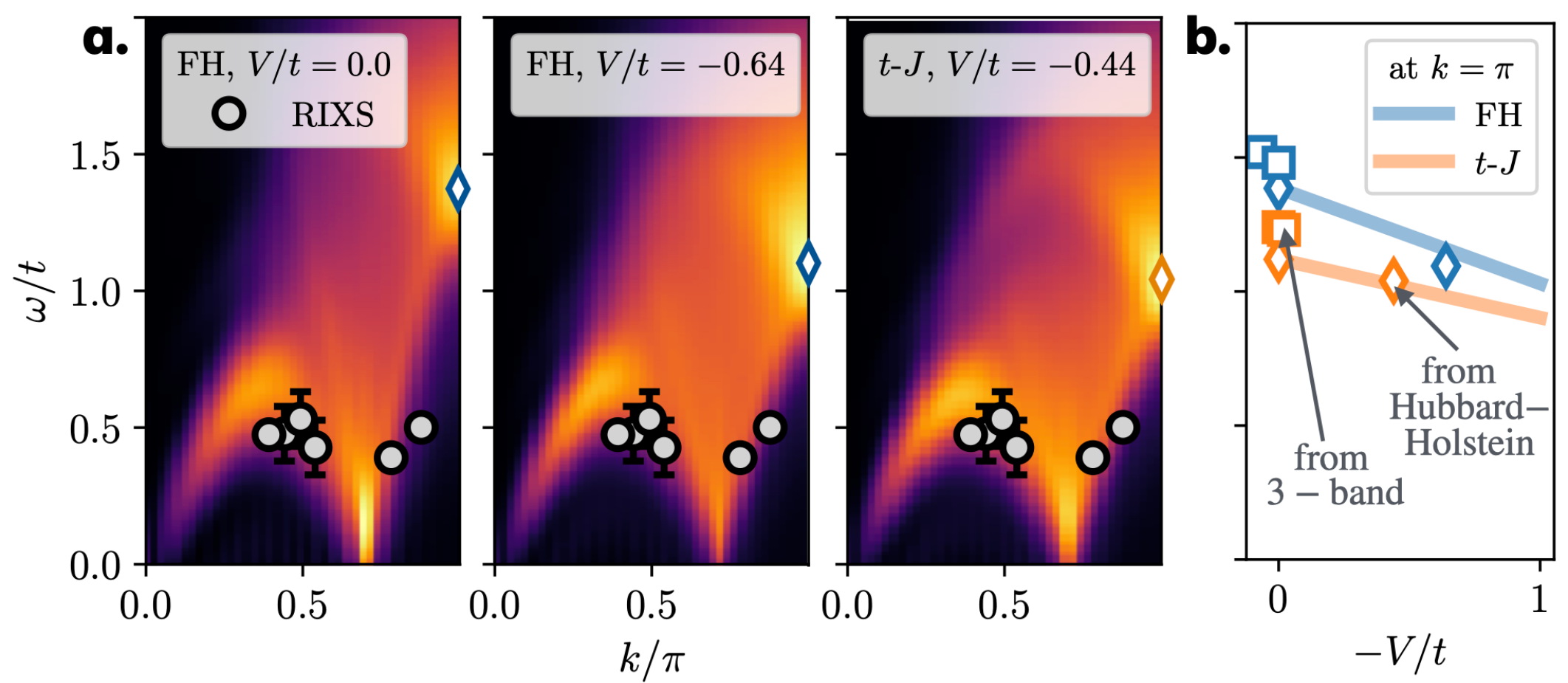}
\caption{Dynamical spin structure factor (DSF) for 1D compounds at $29\%$ doping: \textbf{a.} We compare RIXS measurements from Ref.~\cite{Li2021} (circles) to the DSF for the Fermi-Hubbard model with $V/t=0.0$, $V/t=-0.64$ and the $t$-$J$ model with $V/t=-0.44$ (left to right color plots). Panel \textbf{b} shows how the energy of the peak at $k=\pi$ shifts downwards with $- V/t$ for both models. Squares represent the reconstruction from the 3-band model, diamonds from the Hubbard-Holstein model, see Tab. \ref{tab:Params}; lines are additional results with $V^\prime=0$ for reference. }
\label{fig:1DDSF}
\end{figure}

We now test the reconstructed single-band models -- derived from the ground states -- by evaluating their performance on the DSF for hole doping $\delta=29\%$. In the undoped case, the system exhibits a characteristic two-spinon continuum, analogous to that observed in ladder systems~\cite{Cloizeaux1962}. Upon doping, the gap-closing point shifts to $2k_F$, with $k_F$ the Fermi momentum, and a spin gap opens at $k=\pi$. This is observed for $\delta=29\%$ in Fig.~\ref{fig:1DDSF}\textbf{a} for all considered single-band models, including the bare Fermi-Hubbard model (left). However, a comparison with experimental RIXS measurements~\cite{Li2025} and the theoretical predictions of the bare Fermi-Hubbard model shows a discrepancy of the DSF near the zone boundary: The spectral gap extends to higher energies than observed experimentally. As for the ladders, it was shown that the inclusion of an attraction $V=-t$ can decrease the spectral gap~\cite{Tohyama2024,Tang2024,Shen2024}. 

The Fermi-Hubbard and $t$-$J$ reconstructions both feature attractive interactions, with again a smaller reconstructed  $\vert V\vert $ for the $t$-$J$ model than for the Fermi Hubbard model, resulting in a similar trend for the DSF gap at $k=\pi$, see Fig.~\ref{fig:1DDSF}\textbf{a} (middle and right) and diamonds in Fig.~\ref{fig:1DDSF}\textbf{b}. The reconstruction from the three-band model does not result in such a large attractive term, which in turn gives rise to a larger spectral gap, see squares in Fig.~\ref{fig:1DDSF}\textbf{b}. In general, the gap is significantly lower for the $t$-$J$ model. In particular, the gap of the $t$-$J$ reconstruction with $V/t=-0.44$ is very close to the Fermi-Hubbard model with $V=-t$ that is considered in Refs.~\cite{Tohyama2024,Tang2024,Shen2024}. 

As discussed in Ref.~\cite{Tang2024}, additionally a softening of the spectral peak at $k=\pi$ and decreasing spectral weight at $2k_F$ is observed in both models when increasing $\vert V\vert$, which is in agreement with the theoretical DSF from the Hubbard-Holstein model. Again, we conclude that both Fermi-Hubbard and $t$-$J$ models with similar attraction are in good agreement with the DSF from the experiment.

Note that similar results have been reported in angle-resolved ARPES studies~\cite{Tohyama2024,Tang2023,Feiguin2023}; however, the differences in the intensity of the holon-folding branch are much less pronounced (see SM).\\

\textit{Two-dimensional systems.--} Finally, we turn to the two-dimensional case. To this end, we compare charge and spin correlation functions for cylinders of size $L_x=12$, $L_y=6$ upon doping the half-filled system with two holes. More specifically, we consider the connected density-density correlator for the holes
\begin{align}
    C_{\hat{n}^h}^{c}(\mathbf{i}, \mathbf{j}) = \langle \hat{n}^h_{\mathbf{i}} \hat{n}^h_{\mathbf{j}} \rangle 
- \frac{\langle \hat{N}^h \rangle - 1}{\langle \hat{N}^h \rangle} 
\langle \hat{n}^h_{\mathbf{i}} \rangle \langle \hat{n}^h_{\mathbf{j}} \rangle,
\end{align}
with the hole number operator $\hat{n}^h_{\mathbf{i}}=(1-\hat{n}^\uparrow_{\mathbf{i}})(1-\hat{n}^\downarrow_{\mathbf{i}})$ and $\hat{N}^h=\sum_{\mathbf{i}}\hat{n}^h_{\mathbf{i}}$, and the normalization factor accounting for the finite number of holes in the system. We average this quantity over different reference sites along the periodic direction and, for the Fermi-Hubbard model, correct for doublon-hole fluctuations by subtracting the correlation function obtained from the ground state with a single dopant hole~\cite{Blatz2024}.  Furthermore, we evaluate the staggered spin correlation function, defined as
\begin{align}
    C_{\hat{S}}(\mathbf{i}) = (-1)^{\mathbf{i}_x+\mathbf{i}_y} \langle \hat{S}^{z}_{(0, 2)} \, \hat{S}^{z}_{\mathbf{i}}\rangle,
\end{align}
with the reference site at position $(0,2)$ on the edge of the cylinder.\\

\begin{figure}[t]
\centering
\includegraphics[width=0.49\textwidth]{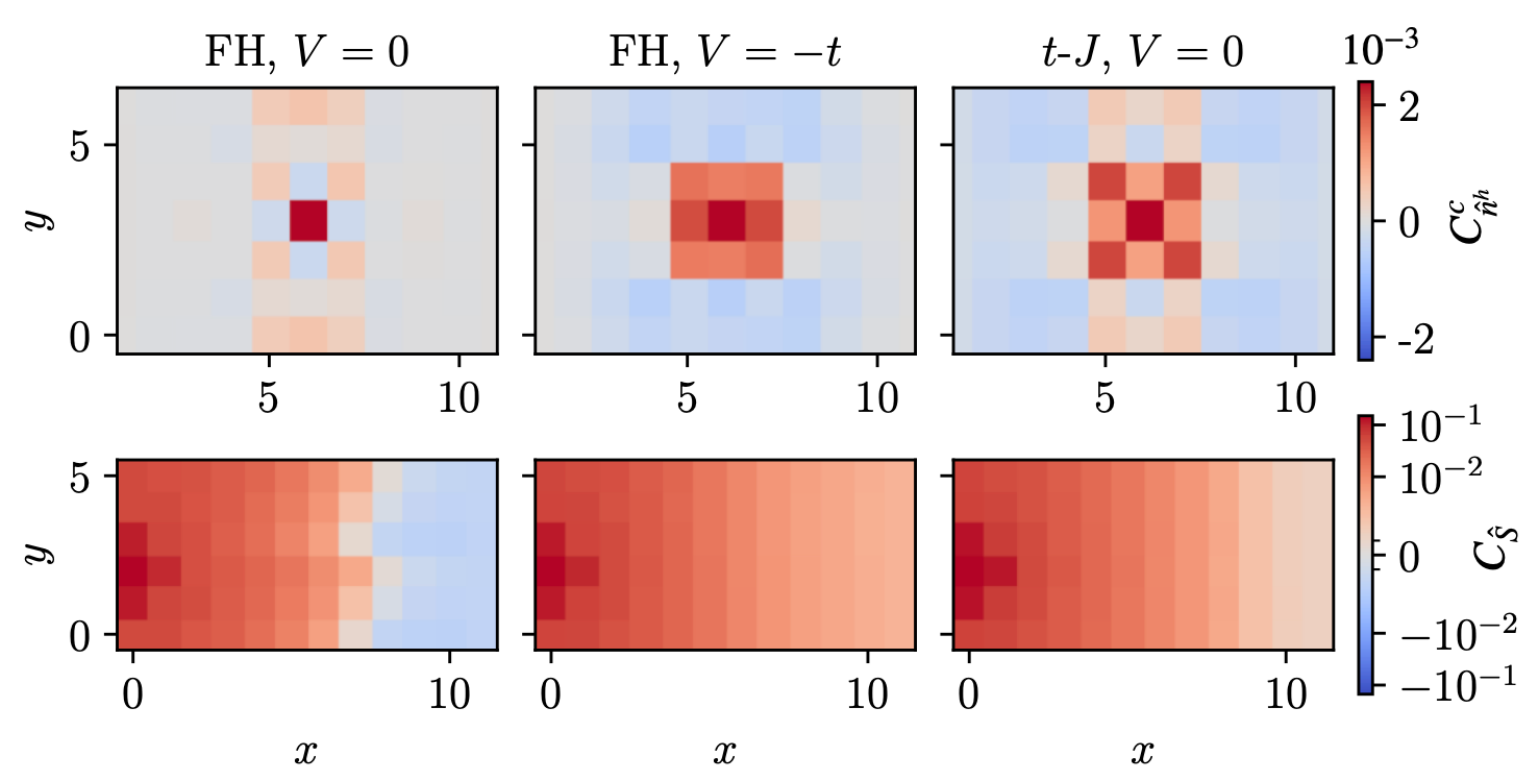}
\caption{Connected charge (top) and staggered spin (bottom) correlations for the Fermi-Hubbard model with $U/t=8.0$ and $V/t=0.0,-1.0$ (left, middle) and the $t$-$J$ model with $J/t=0.5$ and $V/t=0.0$ (right) on cylinders (periodic boundaries in $y$ direction) with $L_x=12$, $L_y=6$ upon doping the half-filled system with two holes. For the Fermi-Hubbard model, the correlations are corrected for doublon-hole fluctuations. }
\label{fig:corrs}
\end{figure}

As shown in the top panel of Fig.~\ref{fig:corrs}, the charge correlations for the bare Fermi-Hubbard model exhibit anticorrelation for nearest neighbors and attractive features for next-nearest neighbors as well as along the periodic direction of the cylinder, indicating the formation of a stripe in the finite size system. This is supported by a sign change in $C_{\hat{S}}$ (bottom panel), consistent with the presence of a spin domain wall that accompanies the stripe. In contrast, for attractive interactions ($V = -t$), the stripe signal vanishes: strong pairing signatures appear in $C_{\hat{n}^h}^{c}$ for both nearest and next-nearest neighbors, and the antiferromagnetic spin correlations extend across the entire system, without any domain wall. This behavior coincides with an increase in the binding energy from $E_B^\mathrm{FH}(V=0)/t = 0.10$ to $E_B^\mathrm{FH}(V = -t)/t = 0.27$.

In the bare $t$–$J$ model, similar pairing features are observed, along with an additional positive contribution from the stripe configuration, reminiscent of the behavior in the bare Fermi-Hubbard model. The antiferromagnetic spin correlations again span the entire system, as in the FH$+V$ case. The binding energy in this case, $E_B^{tJ}(V=0)/t = 0.23$, is comparable to that of the Fermi-Hubbard model with attractive interactions.\\

\textit{Conclusions.--} In this work, we address the question of which minimal single-band model best captures the essential physics of cuprate superconductors, with a particular focus on extended versions of single-band Fermi-Hubbard and $t$-$J$ models. We evaluate their suitability based on the magnetic excitation spectrum probed by the dynamical spin structure factor (DSF) and systematic ground state reconstruction schemes. 

Our key results are two-fold: First, they suggest that for both single-band models additional attractive terms, as suggested by recent experiments~\cite{Padma2025,Scheie2025,Li2025}, can indeed be induced by phonons, while they do not occur in a reconstruction from the three-band model. 

Second, we show that such a Fermi-Hubbard model with attractive interactions shares many properties with the $t$-$J$ model -- even without an explicit attractive interaction for the latter model -- in chains, ladders and two-dimensional systems. Most prominently, both FH$+V$ and the bare $t$-$J$ model are in agreement with recent RIXS measurements on cuprate ladder compounds, where a significant discrepancy between the experiment and the DSF from the bare Fermi-Hubbard model was found~\cite{Scheie2025,Padma2025}. 
This can be attributed to the presence of strongly bound rung pairs in both FH$+V$ and bare $t$-$J$ models. 

Our results highlight both the need for further studies incorporating additional terms in single-band models and the remarkable agreement of the bare $t$-$J$ model with experiments on cuprates resulting from the intrinsic strong pairing in this model.
\\

\textbf{Acknowledgements.} We thank Fabian Grusdt, Johannes Hofmann, Simon Linsel, Matteo Mitrano, Hari Padma, Henning Schlömer, Yao Wang, Krzysztof Wohlfeld and Reja Wilke for valuable discussions. Special thanks to Hari Padma and Matteo Mitrano for providing the RIXS data shown in Fig.~\ref{fig:LadderDSF}. HL acknowledges support by the International Max Planck Research School for Quantum Science and Technology (IMPRS-QST). We acknowledge funding from the Deutsche Forschungsgemeinschaft (DFG, German Research Foundation) under Germany's Excellence
Strategy—EXC2111—390814868. Numerical simulations were performed on the Arnold Sommerfeld Center for Theoretical Physics computing cluster.\\
%

\bibliographystyle{apsrev4-1}
\bibliography{main}

\newpage
\onecolumngrid
\widetext

\appendix


\begin{center}
    \vspace{2em}
    {\large\bfseries Supplemental Material 
    \par}
    \vspace{2em}
\end{center}

\section{Reconstruction Schemes for Effective Hamiltonians}

In this section, we describe the reconstruction procedures used to derive effective Hamiltonians from the Hubbard–Holstein chains, three-band Hubbard models and the Fermi-Hubbard ladder considered in this work.

\subsection{Reconstruction from the Hubbard–Holstein Model \label{appendix:HamRec1}}

\subsubsection{Reconstruction scheme}
To reconstruct effective single-band Fermi–Hubbard and $t$–$J$ Hamiltonians, we follow the ground-state reconstruction approach outlined in Ref.~\cite{Qi2019}. This method involves reconstructing the effective Hamiltonian using measurements performed on the ground state $v$. Here, we closely follow the explanation presented in Ref.~\cite{Qi2019}.

The goal of the reconstruction is to find the operator $\hat{\mathcal{H}}_\mathrm{eff}$ for which the variance w.r.t. the ground state of the original Hamiltonian $v$ is reduced, i.e.
\begin{align}
    \langle \hat{\mathcal{H}}_\mathrm{eff}^2\rangle_v-\langle \hat{\mathcal{H}}_\mathrm{eff}\rangle_v^2=\epsilon
    \label{eq:variancecond}
\end{align}
where $\langle \cdot \rangle_v$ denotes the expectation value in the state $v$, and $\epsilon $ is small. To do that, we construct $ \hat{\mathcal{H}}_\mathrm{eff}$ from a set of $\{L_i\}$ that span the space of the effective Hamiltonian, i.e. $ \hat{\mathcal{H}}_\mathrm{eff}=\sum_i w_i\hat{L}_i$. The key object in this reconstruction is then given by the correlation matrix $M^{(v)}$, defined as:
\begin{equation*}
M^{(v)}_{ij} = \frac{1}{2}\langle\{L_i, L_j\} \rangle_v -\langle L_i \rangle_v \langle L_j \rangle_v.
\end{equation*}
The condition \eqref{eq:variancecond} above can then be reformulated as an eigenvalue problem, namely as
\begin{align}
    w^TM^{(v)}w=\epsilon.
\end{align}

An effective Hamiltonian can be reconstructed if the eigenvalue spectrum of $M^{(v)}$,
\begin{equation}
\text{spec}(M^{(v)}) = \{\lambda^{(v)}_i\}_{i=1}^N,
\end{equation}
with $\lambda_i^{(v)}$ ordered from the lowest to the highest eigenvalue, satisfies $\lambda^{(v)}_2 > 0$. Specifically, if $\lambda^{(v)}_1 = 0$, then the effective Hamiltonian is uniquely determined. The corresponding eigenvector encodes the reconstructed Hamiltonian parameters. In our setting, where the reconstruction is performed in a different Hilbert space than the original, exact recovery is not possible, and we find that all $\lambda^{(v)}_2 > 0$. In this case, the smallest eigenvalue $\lambda^{(v)}_1$ serves as a measure of the reconstruction error.\\

More precisely, we use the translationally invariant formulation of this scheme. We denote the Fourier-transformed operator basis as $\{\widetilde L_{q\alpha}\}$, where $q$ is the momentum. Specifically, we define:
\begin{equation}
\widetilde L_{q\alpha} = \frac{1}{\sqrt{n}} \sum_{x=1}^n L_{x\alpha} e^{iqx},
\end{equation}
and introduce the translationally invariant correlation matrix:
\begin{align}
\widetilde M^{(v)}_{p\alpha, q\beta} = \frac{1}{2} \langle 
\{ \widetilde L_{p\alpha}, \widetilde L_{q\beta} \} \rangle_v - \langle \widetilde L_{q\alpha} \rangle_v \langle \widetilde L_{p\beta} \rangle_v.
\end{align}
In this formulation, only the zero-momentum sector of the correlation spectrum is relevant for reconstructing a translationally invariant Hamiltonian from its eigenstate.\\

For the reconstruction of the effective single-band Fermi–Hubbard and $t$–$J$ Hamiltonians from the Hubbard-Holstein model, we simulate the Hubbard-Holstein model with $L=12,24,36$ sites with $2,4,6$ holes respectively and reconstruct the effective single-band Fermi–Hubbard and $t$–$J$ Hamiltonians for all system sizes and extrapolate to $L\to \infty$. In general, the reconstructed parameters do not strongly depend on $L$. Details on the simulations of the Hubbard-Holstein model can be found in Sec.~\ref{appendix:MPS}.

\subsubsection{Reconstruction results to the Fermi-Hubbard model}
The reconstructed parameters for the effective single-band Hubbard model are displayed in Table~\ref{tab:HHtoFH}.
\begin{table}[htp]
\begin{tabular}{|c|c|c|c|c|c|c|}
\hline
system size                                & model type & $U/t$ & $V/t$ &$V^\prime/t$ & $t^\prime/t$ & error \\ \hline
                                           &  FH &  8.31 & - & - & - & 0.0012        \\
                                           & extended FH & 7.02 & -1.01 & -0.59&-& 0.0004      \\
\multirow{-3}{*}{$L=12$}                   & extended FH$^\prime$  & 6.99 & -1.0 & -0.53& 0.05 & 0.0003       \\ \hline
                                           & FH &  8.18 & - & -  & - & 0.0007      \\
                                           &extended FH &  7.19 & -0.85 &-0.47 & - & 0.0004       \\
\multirow{-3}{*}{$L=24$}                   &extended FH$^\prime$  & 7.12 & -0.86&-0.42 & 0.05 & 0.0004     \\ \hline
                                           &FH &  8.14  & - & - & - & 0.0006        \\
                                           &extended FH &  7.29 & -0.75 & -0.42& - & 0.0004     \\
\multirow{-3}{*}{$L=36$}                   &extended FH$^\prime$ &  7.2 & -0.77 &-0.37& 0.05 & 0.0004       \\ \hline
\rowcolor[HTML]{EFEFEF} 
\cellcolor[HTML]{EFEFEF}                   &   FH &  8.05 & - & - &  - &       \\
\rowcolor[HTML]{EFEFEF} 
\cellcolor[HTML]{EFEFEF}                   &  extended FH & 7.41 & -0.64&-0.34&-&        \\
\rowcolor[HTML]{EFEFEF} 
\multirow{-3}{*}{\cellcolor[HTML]{EFEFEF}$L\to \infty$} &  extended FH$^\prime$ &  7.29 & -0.67 & -0.3&0.07 &       \\ \hline
\end{tabular}
\caption{Reconstructed parameters (in units of the effective reconstructed hopping $t$) for the single-band Fermi-Hubbard (FH) model from the Hubbard-Holstein model for system sizes $L=12,24,36$ and extrapolated to $L\to \infty$. Extended FH$^{(\prime)}$ denotes FH with a nearest (and next-nearest, distance 2) neighbor density interaction $V^{(\prime)}$ (and a next-nearest neighbor (distance 2) hopping term $t^\prime$). The errors are given w.r.t. the hopping amplitude of the original model. \label{tab:HHtoFH}}
\end{table}

\subsubsection{Reconstruction results to the $t$-$J$ model}
In order to reconstruct the effective $t$-$J$ model, we perform the same reconstruction starting from the single-band Fermi-Hubbard model in the previous sub-section and reconstruct the effective $t$-$J$ description. The reconstructed parameters for the effective single-band $t$-$J$ model are displayed in Table~\ref{tab:HHtotJ}. When including a next-nearest neighbor $V^\prime$, the sign of $t_3$ changes and $J$ becomes larger than expected. Since the reconstructed $V^\prime$ is anyways vanishingly small, we neglect it in the main text and focus on the reconstruction parameter labeled by $t$-$J$ and extended $t$-$J$.

\begin{table}[htp]
\begin{tabular}{|c|c|c|c|c|c|c|c|c|}
\hline
system size                                & model type & $J/t$ & $t_3/t$ & $t^\prime/t$ &$J^\prime/t$ &$V/t$&$V^\prime/t$& error \\ \hline
                                           &  $t$-$J$ & 0.47 & 0.07  & - & - & -& -&  0.00025       \\
                                           & extended $t$-$J$  & 0.50 & 0.07 & -& -&-1.08& -  & 0.00012      \\
\multirow{2}{*}{$L=12$}                   & extended $t^\prime$-$J^\prime$& 0.50  & 0.07 & 0.0 & -0.07 & -0.99 & -& 0.00014 \\
&extended $t$-$J$ with $V^\prime$& 0.73  & -0.04& -& - & -0.57 & 0.03 & 0.00177\\
& extended $t$-$J^\prime$ with $V^\prime$& 0.53 & -0.11 & 0.03 & -0.2 & -0.67 & 0.0 & 0.00056

\\ \hline
                                           & $t$-$J$ & 0.51 &  0.06  & - & - & -& -& 0.00023      \\
                                           &extended $t$-$J$ & 0.5 & 0.06 &  -& -&-0.78 & -& 0.00007       \\
\multirow{2}{*}{$L=24$}                   &extended $t^\prime$-$J^\prime$& 0.51 & 0.04 & 0.0 & -0.09 & -0.78 & -&0.00010    \\
&extended $t$-$J$ with $V^\prime$& 0.71  & -0.02&  -& - & -0.59 & -0.01 & 0.00196\\
&extended $t$-$J^\prime$ with $V^\prime$& 0.66  & -0.09 & 0.05 & -0.15 & -0.59 & 0.0 & 0.00096\\
\hline
                                           &$t$-$J$ & 0.51 &  0.06 & - & - & -& -& 0.00023       \\
                                           &extended $t$-$J$ & 0.5  & 0.06 &  -& -&-0.64 & -& 0.00007     \\
\multirow{2}{*}{$L=36$}                   &extended $t^\prime$-$J^\prime$& 0.51  & 0.03 & -0.01 & -0.09 & -0.66 & -& 0.00007      \\ 
&extended $t$-$J$ with $V^\prime$& 0.7  & -0.02 & -& -& -0.59 & -0.01 & 0.00199 \\
&extended $t$-$J^\prime$ with $V^\prime$& 0.68 &  -0.08 & 0.05 & -0.14 & -0.57 & 0.0 & 0.00104\\

\hline
\rowcolor[HTML]{EFEFEF} 
\cellcolor[HTML]{EFEFEF}                   &   $t$-$J$ & 0.51 &  0.06 & - & - & -& -&         \\
\rowcolor[HTML]{EFEFEF} 
\cellcolor[HTML]{EFEFEF}                   &  extended $t$-$J$ & 0.52  & 0.05 &  -& -&-0.44 & -&     \\
\rowcolor[HTML]{EFEFEF} 
\multirow{-2}{*}{\cellcolor[HTML]{EFEFEF}$L\to \infty$} & extended $t^\prime$-$J^\prime$& 0.51  & 0.01 & -0.01 & -0.11 & -0.51 & -&       \\ 
\rowcolor[HTML]{EFEFEF} 
&extended $t$-$J$ with $V^\prime$& 0.69  & -0.01 & -& -& -0.6 & -0.04 &  \\
\rowcolor[HTML]{EFEFEF} 
\cellcolor[HTML]{EFEFEF}&extended $t$-$J^\prime$ with $V^\prime$& 0.76 &  -0.06 & 0.06 & -0.11 & -0.52 & -0.01 & \\
\hline
\end{tabular}
\caption{Reconstructed parameters (in units of the effective reconstructed hopping $t$) for the single-band $t$-$J$ model from the Hubbard-Holstein model for system sizes $L=12,24,36$ and extrapolated to $L\to \infty$. Extended $t$-$J$$^{(\prime)}$ denotes $t$-$J$ with a nearest-neighbor density interaction $V$ (and next-nearest neighbor $t^\prime$ and $J^\prime$ terms at distance $2$). The errors are given w.r.t. the hopping amplitude of the original model.}
\label{tab:HHtotJ}
\end{table}

\begin{figure}[t]
\centering
\includegraphics[width=0.95\textwidth]{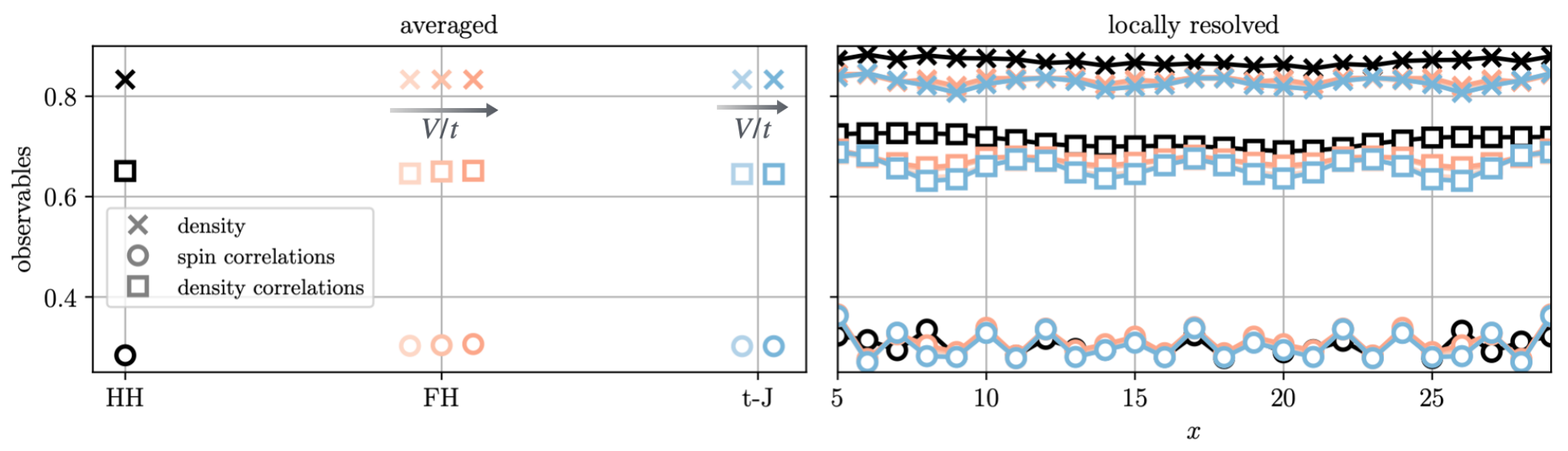}
\caption{Averaged (left) and locally resolved (right) observables for the Hubbard-Holstein model (HH) and the reconstructed models. The reconstructed models are the Fermi-Hubbard model (FH), including the usual form of FH, extended FH and extended FH with nearest neighbor hopping (left to right), as well as reconstructed $t$-$J$ (without (left) and with (right) nearest neighbor density interaction.) }
\label{fig:recontructed_obs}
\end{figure}

Exemplary observables for the Hubbard-Holstein model (HH) and the reconstructed models are shown in Fig.~\ref{fig:recontructed_obs}.

\subsection{Reconstruction from the three-band model \label{appendix:HamRec2}}
\subsubsection{Reconstruction scheme}
In order to reconstruct a single-band model from the three-band Fermi-Hubbard model, we use the reconstruction schemes from Refs.~\cite{Jiang2021,Sawaya2022}. Here, we closely follow Ref.~\cite{Jiang2021}. 

We start from the three-band model with two oxygen $p_x$ and $p_y$ sites and one copper $d$ site. The goal is to define Wannier functions (WF) centered on the copper sites and construct a matrix $A$ out of them, defining a single-particle transformation from the three-band basis $\{C^\dagger\}=\{d^\dagger,p_x^\dagger,p_y^\dagger\}$ to the single-band (or WF) basis $\{c^\dagger\}$, $c^\dagger_j=\sum_j A_{ji}C^\dagger_i$.
Following Ref.~\cite{Jiang2021}, we define the correlation matrix
\begin{align}
    M_{\alpha \beta}=\sum_{\sigma}\langle C^\dagger_{\alpha\sigma} C_{\beta\sigma} \rangle,
\end{align}
with $C^\dagger=\{d^\dagger,p_x^\dagger,p_y^\dagger\}$. Diagonalization of this matrix yields a basis of Wannier functions or natural orbitals $\{\psi_i\}$, but they are not necessarily centered at the copper sites. These $\{\psi_i\}$ can be centered around the copper sites by superposing them as follows:
\begin{equation}
    \psi^C_j(\Vec{r})=\sum_i \psi_i(\Vec{r}=\Vec{r}_j) \psi_i(\Vec{r})
\end{equation}
where $\Vec{r}_j$ is the position vector corresponding to Cu site $j$. In order to orthonormalize $\{\psi^C_j(\Vec{r})\}$ by a minimal rotation denoted by $O_{jj'}=\langle \psi^C_{j}(\Vec{r})~|~\psi^C_{j'}(\Vec{r}) \rangle$, yielding
\begin{equation}
\begin{split}
\psi^{CO}_j(\Vec{r})=\sum_j {[O^{-\frac{1}{2}}]}_{jj'}~\psi^C_{j'}(\Vec{r}_i).
\end{split}
\end{equation}
$A_{ji}$ is constructed out of all $\psi^{CO}_j(\Vec{r})$ for all $j$ copper sites.

As in Ref.~\cite{Jiang2021}, we can read off the effective Hamiltonian terms from products of the $A$-matrix. In particular, we consider up to nearest-neighbor (NN) interactions, specifically

\begin{enumerate}
    \item a NN hopping defined by
\begin{align}
    t_{\alpha\beta}\sum_i A_{i,\alpha}A_{i+1,\beta}\hat{c}^\dagger_{i,\sigma}\hat{c}_{i+1,\sigma},
\end{align}
\item on-site repulsion defined by
\begin{align}
    U_{\alpha}\sum_i A_{i,\alpha}^4\hat{n}^\dagger_{i,\uparrow}\hat{n}_{i\downarrow},
\end{align}
\item a NN density assisted tunneling
\begin{align}
    U_{\alpha}\sum_i A_{i,\alpha}A_{i+1,\alpha}A_{i+1,\alpha}A_{i+1,\alpha}\hat{c}^\dagger_{i,\sigma}\hat{c}_{i+1,\sigma}\hat{n}_{i+1,\bar{\sigma}},
\end{align}
\item a NN density interaction
\begin{align}
    U_{\alpha}\sum_i A_{i,\alpha}A_{i,\alpha}A_{i+1,\alpha}A_{i+1,\alpha}\hat{n}_{i,\sigma}\hat{n}_{i+1,\bar{\sigma}},
\end{align}
and the terms with the same amplitude, corresponding to correlated hopping processes $\hat{c}^\dagger_{i,\sigma}\hat{c}_{i+1,\sigma}\hat{c}^\dagger_{i,\bar{\sigma}}\hat{c}_{i+1,\bar{\sigma}}$ and $\hat{c}^\dagger_{i,\sigma}\hat{c}_{i+1,\sigma}\hat{c}^\dagger_{i+1,\bar{\sigma}}\hat{c}_{i,\bar{\sigma}}$. These terms are not considered in Ref.~\cite{Jiang2021}, but partly in Ref.~\cite{Sawaya2022}.
\end{enumerate}

In analogy to Ref.~\cite{Sawaya2022}, in most cases we would like to obtain an effective Hubbard model with a density interaction term $V$ from the Hamiltonian including terms 1.-4.. However, none of 1.-4. gives rise to a full $V$ term. Since 4. contains density interactions for opposite spins as well as correlated hopping terms that potentially give rise to pairing, we apply the ground state reconstruction scheme from Sec.~\ref{appendix:HamRec1} on top of the Wannier reconstruction to determine if these terms give rise to an effective $V$ term.\\

We use the three-band parameters from Ref.~\cite{Jiang2021}, namely $t_{pd}=1.0$, and take $t_{pp}=0.5$, $U_d=6.0$, $U_p=3.0$, and $\Delta_{pd}=3.5$.

\subsubsection{Reconstruction results to the Fermi-Hubbard model}
The reconstructed parameters for the effective single-band Hubbard model are displayed in Table~\ref{tab:LiebtoFH}. More precisely, the procedure that we apply is as follows:
\begin{enumerate}
    \item We downfold from the three-band model to the single-band Fermi Hubbard model including terms 2a 1.-4. using the Wannier downfolding in Sec.~\ref{appendix:HamRec2}. The respective parameters are labelled Jiang et al.~\cite{Jiang2021} in the table.
    \item As explained in the previous section, we would like to obtain an effective Hubbard model with a density interaction term $V$ from the Hamiltonian including terms 1.-4.. However, none of 1.-4. gives rise to a full $V$ term. Since 4. contains density interactions for opposite spins as well as correlated hopping terms that potentially give rise to pairing, we apply the ground state reconstruction scheme from Sec.~\ref{appendix:HamRec1} on top of the Wannier reconstruction to determine if these terms give rise to an effective $V$ term. The respective errors are given in the last column.
\end{enumerate}
Note that we neglect all next-nearest neighbor hopping contributions here, since in our one dimensional setting these terms have distance 2 and consequently are very small, see also Ref.~\cite{Jiang2021}.

\begin{table}[htp]
\begin{tabular}{|c|c|c|c|c|c|}
\hline
system size                                & model type & $U/t$ & $t_n/t$& $V/t$ or terms of same order &  error \\ \hline
                                           & FH as in Jiang et al. &  11.32 & 0.64 &  0.13  & -      \\
                                           & FH &  11.29 & 0.64 & - & 0.000017   \\
                        \multirow{-4}{*}{$L=12$}                    & extended FH & 11.37 & 0.64 & 0.09 &  0.000015      \\
\hline
                                           & FH as in Jiang et al.  &  11.36 & 0.64 &  0.13 & -  \\
                                           & FH &  11.29 & 0.64  & - & 0.000017   \\
                   \multirow{-3}{*}{$L=24$}                          &extended FH &  11.39 & 0.65 & 0.08 & 0.000016       \\
\hline
                                           & FH as in Jiang et al.  &  11.39 & 0.64 &  0.13 & -  \\
                                           & FH &  11.30 & 0.64  & - & 0.000017   \\
                   \multirow{-3}{*}{$L=36$}                          &extended FH &  11.4 & 0.65 & 0.08 & 0.000016       \\
\hline
\rowcolor[HTML]{EFEFEF} 
\cellcolor[HTML]{EFEFEF}                   &  FH as in Jiang et al. &  11.39 & 0.64 &  0.13  &      \\
\rowcolor[HTML]{EFEFEF} 
\cellcolor[HTML]{EFEFEF}                   &  FH &  11.32 & 0.64 &  -  &      \\
\rowcolor[HTML]{EFEFEF} 
\rowcolor[HTML]{EFEFEF} 
\multirow{-3}{*}{\cellcolor[HTML]{EFEFEF}$L\to \infty$}                   &  extended FH &  11.41 & 0.65 & 0.07  &       \\
 \hline
\end{tabular}
\caption{Reconstructed parameters (in units of the effective reconstructed hopping $t$) for the single-band Fermi-Hubbard (FH) model from the three-band Hubbard model for system sizes $L=12,24,36$ and extrapolated to $L\to \infty$. The downfolding as in Jiang et al.~\cite{Jiang2021} yields the parameters in the first line of each system size, where the terms listed in 2a4. are labeled by \enquote{$V/t$ or terms of same order}. FH$^{(\prime)}$ denotes FH with a nearest-neighbor density interaction $V$ 
. The errors are given w.r.t. the hopping amplitude of the original model. }
\label{tab:LiebtoFH}
\end{table}

\subsubsection{Reconstruction results to the $t$-$J$ model}
In order to reconstruct the effective $t$-$J$ model, we perform the same reconstruction starting from the effective single-band model in the previous sub-section. The reconstructed parameters for the effective single-band $t$-$J$ model are displayed in Table~\ref{tab:LiebtotJ}.

\begin{table}[htp]
\begin{tabular}{|c|c|c|c|c|c|}
\hline
system size                                & model type & $J/t$ & $t_3/t$ & $V/t$& error \\ \hline
                                           &  $t$-$J$ & 0.69 & 0.07  & -&  0.0002       \\
\multirow{-2}{*}{$L=12$}                   & extended $t$-$J$& 0.63  & 0.08 &  0.0 & 0.0001     \\ \hline
                                           &$t$-$J$ & 0.86&  0.07 & -& 0.0002       \\
\multirow{-2}{*}{$L=24$}                   &extended $t$-$J$& 0.65  & 0.08  & -0.01 & 0.0001      \\ \hline
                                           &$t$-$J$ & 0.86&  0.07 & -& 0.0002       \\
\multirow{-2}{*}{$L=36$}                   &extended $t$-$J$& 0.65  & 0.08  & -0.01 & 0.0001      \\ \hline
\rowcolor[HTML]{EFEFEF} 
\cellcolor[HTML]{EFEFEF}                   &   $t$-$J$ & 0.68 &  0.08 & -&       \\
\rowcolor[HTML]{EFEFEF} 
\multirow{-2}{*}{\cellcolor[HTML]{EFEFEF}$L\to \infty$} &extended $t$-$J$& 0.66  & 0.08  & -0.02 &       \\ \hline
\end{tabular}
\caption{Reconstructed parameters (in units of the effective reconstructed hopping $t$) for the single-band $t$-$J$ model from the three-band Hubbard model for system sizes $L=12,24,36$ and extrapolated to $L\to \infty$. Extended $t$-$J^{(\prime)}$ denotes $t$-$J$ with a nearest-neighbor density interaction $V$ (and next-nearest neighbor hopping $t^\prime$ and spin exchange $J^\prime$ terms at distance 2). The errors are given w.r.t. the hopping amplitude of the original model. }
\label{tab:LiebtotJ}
\end{table}

\subsection{Reconstruction from the Fermi-Hubbard ladder}

To derive an effective $t$-$J$ ladder model from the extended Fermi-Hubbard ladder, we apply the reconstruction scheme from Sec.~\ref{appendix:HamRec1} to the extended Fermi-Hubbard ladder from Ref.~\cite{Padma2025},
\begin{align}
    H = & -t \sum_{i,l,\sigma} \left( \hat{c}_{i,l,\sigma}^\dagger \hat{c}_{i+1,l,\sigma} + \text{h.c.} \right)
- t_\perp \sum_{i,\sigma} \left( \hat{c}_{i,1,\sigma}^\dagger \hat{c}_{i,2,\sigma} + \text{h.c.} \right) - t_\mathrm{diag} \sum_{i,\sigma} \left( \hat{c}_{i,1,\sigma}^\dagger \hat{c}_{i+1,2,\sigma} + \hat{c}_{i,2,\sigma}^\dagger \hat{c}_{i+1,1,\sigma} + \text{h.c.} \right) \notag\\
& + U \sum_{i,l} \hat{n}_{i,l,\uparrow} \hat{n}_{i,l,\downarrow}
+ V \sum_{\langle (i,l),(i',l') \rangle,\sigma,\sigma'} \hat{n}_{i,l,\sigma} \hat{n}_{i',l',\sigma'}
\end{align}
namely with nearest-neighbor density attraction amplitude $V_\parallel/t_\parallel=V_\perp/t_\parallel=-1.25$, perpendicular hopping $t_\perp/t_\parallel=0.84$, on-site repulsion $U/t_\parallel=8.0$ or $U/t_\parallel=8.0+V$ and a diagonal hopping $t_\mathrm{diag}/t_\parallel=-0.3$.

\begin{table}[htp]
\begin{tabular}{|c|c|c|c|c|c|c|c|c|c|c|}
\hline
system size                                & model type & $J_\parallel/t_\parallel$ & $t_3^\parallel/t$& $t_\perp/t_\parallel$ &  $J_\perp/t_\parallel$ & $t_3^\perp/t_\parallel$ & $V_\parallel/t_\parallel=V_\perp/t_\parallel$&$t_\mathrm{diag}/t_\parallel$&$J_\mathrm{diag}/t_\parallel$ & error \\ \hline
$L=33$& extended $t$-$J$ from FH$+V$ with $U/t_\parallel=8.00$ & 0.38 & 0.08 & 0.83& 0.20& 0.04 &-0.98&-0.15&-0.01& 0.002      \\ \hline
$L=33$& extended $t$-$J$ from FH$+V$ with $U/t_\parallel=6.75$ & 0.36 & 0.08 & 0.81& 0.17& 0.04 &-1.07&-0.13&-0.02& 0.002      \\ \hline
\end{tabular}
\caption{Reconstructed parameters (in units of the effective reconstructed hopping $t_\parallel$) for the single-band extended Fermi-Hubbard ladder from Ref.~\cite{Padma2025} to an extended $t$-$J$ ladder for system size $L=33$. Extended $t$-$J$ denotes a $t$-$J$ ladder with a nearest-neighbor density interaction $V_\perp=V_\parallel$. The errors are given w.r.t. the $t_\parallel$ hopping amplitude of the original model. }
\label{tab:ladder}
\end{table}

\section{Comparison with Lang-Firsov transformation}

\begin{figure*}[htp]
\centering
\includegraphics[width=0.99\textwidth]{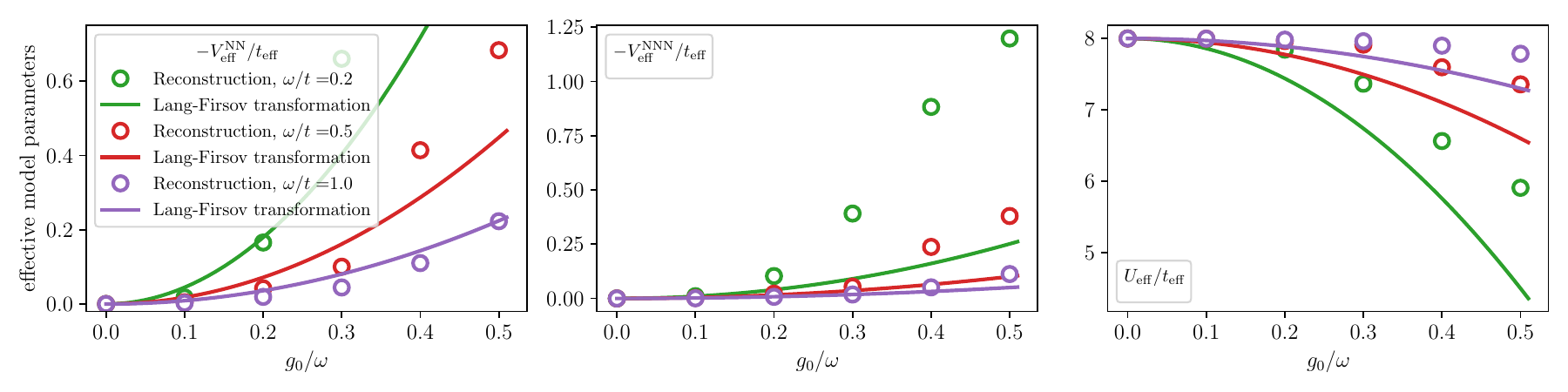}
\caption{Comparison of reconstructed parameters from our reconstruction scheme (markers) and the Lang-Firsov transformation (lines) for different $\omega/t$ indicated by the different colors and $g_1=g_0/\sqrt{5}$. Note that we assume $t_\mathrm{eff}=1$ for the Lang-Firsov transformation results, which is only approximately valid for small $\frac{g_0}{\omega}, \frac{g_1}{\omega}\ll 1$. In this regime, the reconstruction and Lang-Firsov effective model parameters agree well. }
\label{fig:LangFirsov}
\end{figure*}

To evaluate the performance of our reconstruction from the Hubbard-Holstein model \eqref{eq:HH}, we compare the results to a Lang-Firsov transformation of the Hubbard-Holstein Hamiltonian. 

The Lang-Firsov transformation is given by
\begin{align}
    S = \sum_i\hat{n}_i\left( \lambda_0 (\hat{b}^\dagger_i -\hat{b}_i)+\lambda_1 \sum_{\delta=\pm 1}(\hat{b}^\dagger_{i+\delta} -\hat{b}_{i+\delta}) \right) 
    \label{eq:LangFirsov}
\end{align}
with $\lambda_\mu = \frac{g_\mu}{\omega_0}$.
The bosonic operators then transform as follows:
\begin{align}
    \tilde{\hat{b}}_i = e^S \hat{b}_i e^S = \hat{b}_i +[S, \hat{b}_i]+\frac{1}{2!}[S,[S,\hat{b}_i]]+\dots =\lambda_0\hat{n}_i -\lambda_1 (\hat{n}_{i+1}+\hat{n}_{i-1})
\end{align}
since $[S, \hat{b}_i]=\lambda_0 [\hat{n}_i(\hat{b}^\dagger_i -\hat{b}_i), \hat{b}_i]+\lambda_1[\hat{n}_i(\hat{b}^\dagger_{i\pm 1} -\hat{b}_{i\pm 1}), \hat{b}_i]=\lambda_0 \hat{n}_i[(\hat{b}^\dagger_i -\hat{b}_i), \hat{b}_i]+\hat{n}_i\lambda_1[(\hat{b}^\dagger_{i\pm 1} -\hat{b}_{i\pm 1}), \hat{b}_i]=\lambda_0\hat{n}_i (-1)+\lambda_1 (\hat{n}_{i-1}+\hat{n}_{i-1} (-1)$ and $[S,[S, \hat{b}_i]]=0$.

For the fermionic operators, we find
\begin{align}
    \tilde{\hat{c}}_i = \hat{c}_i \mathrm{exp}\left( -\lambda_0(\hat{b}^\dagger_i -\hat{b}_i)-\lambda_1 (\sum_{\delta=\pm1}\hat{b}^\dagger_{i+\delta} -\hat{b}_{i+\delta})\right)=:\hat{c}_i e^{-\hat{\Lambda}_i}.
    \label{eq:Lambda}
\end{align}
This can be seen from
\begin{align}
    [S, \hat{c}_i]=\lambda_0 [\hat{n}_i(\hat{b}^\dagger_i -\hat{b}_i), \hat{c}_i]+\lambda_1[\hat{n}_i\sum_{\delta=\pm1}(\hat{b}^\dagger_{i+\delta} -\hat{b}_{i+\delta}), \hat{c}_i]= -\lambda_0(\hat{b}^\dagger_i -\hat{b}_i)-\lambda_1 \sum_{\delta=\pm1}(\hat{b}^\dagger_{i+\delta} -\hat{b}_{i+\delta}) 
\end{align}
and higher orders $[S,[S,\hat{c}_i]]$. 

With these transformed operators, we get the following transformed Hamiltonian, in which the phononic and electronic degrees of freedom are decoupled:
\begin{align}
    \hat{\mathcal{H}}_{LF}=&-t\sum_{\langle i,j\rangle}(\hat{c}_i^\dagger \hat{c}_je^{\hat{\Lambda}_i}e^{-\hat{\Lambda}_j}+\mathrm{h.c.})+(U-2\frac{g_0^2}{\omega_0}-4\frac{g_1^2}{\omega_0})\sum_i\hat{n}_{i\uparrow}\hat{n}_{i\downarrow}-\left(\frac{g_0^2}{\omega_0}+2\frac{g_1^2}{\omega_0}\right)\sum_i(\hat{n}_{i\uparrow}^2+\hat{n}_{i\downarrow}^2)\notag \\
    &-\frac{2g_0g_1}{\omega_0}\sum_{\langle i,j \rangle}\hat{n}_i\hat{n}_j-\frac{g_1^2}{\omega_0}\sum_{\langle\langle i,j \rangle\rangle}\hat{n}_i\hat{n}_j,
\end{align}
which includes a nearest and next-nearest neighbor attraction. We define $U\mathrm{eff}^{LF}=U-2\frac{g_0^2}{\omega_0}-4\frac{g_1^2}{\omega_0}$ and $V_\mathrm{eff}^LF=-\frac{2g_0g_1}{\omega_0}$. Also the hopping gets renormalized to 
\begin{align}
    t_\mathrm{eff}=t \frac{1}{N_s-1}\sum_{\langle i,j\rangle}(e^{\hat{\Lambda}_i-\hat{\Lambda}_j}+e^{\hat{\Lambda}_j-\hat{\Lambda}_i})/2= t \frac{1}{N_s - 1} \sum_{\langle i, j \rangle} \left[1 + \frac{1}{2}(\hat{\Lambda}_i - \hat{\Lambda}_j)^2 + \frac{1}{24}(\hat{\Lambda}_i - \hat{\Lambda}_j)^4 + \cdots\right]
\end{align}

In the following comparison, we will assume $t_\mathrm{eff}= t$, i.e. we assume $(\hat{\Lambda}_i - \hat{\Lambda}_j)^{2n}\ll 1$. From Eq. \eqref{eq:Lambda}, one can see that this implies small expectation values of combinations of the phononic operators and $\frac{g_0}{\omega}, \frac{g_1}{\omega}\ll 1$.  
Fig.~\ref{fig:LangFirsov} shows that in this regime, the nearest-neighbor attraction
$V_\mathrm{eff}^\mathrm{NN}$ (left), the next-nearest neighbor attraction $V\mathrm{eff}^\mathrm{NNN}$ (middle) and the on-site repulsion $U_\mathrm{eff}$ (right) agree well.

\section{Matrix-product state calculations \label{appendix:MPS}}
For all matrix-product state (MPS) calculations, we use the package SyTen~\cite{syten1,syten2}.

\subsubsection{Ground states of Fermi-Hubbard and $t$-$J$ models}
In all ground state calculations of the 1D Fermi-Hubbard and $t$-$J$ model presented in the main text, we use $U(1)\times U(1)$ (conservation of particle number and total spin $S^z$, projected to the $z$-direction) and bond dimensions up to $\chi=1024$. For the ladders, we use bond dimensions up to $\chi=1536$. \\
For the two-dimensional Fermi-Hubbard and $t$-$J$ cylinders, we exploit the full $U(1)\times SU(2)$ symmetry corresponding to the conservation of the particle number and the total spin $S$. For even particle numbers, we work in the $S = 0$ sector. The single-hole calculation used to determine the binding energy is performed in the $S=1/2$ sector.
We keep a maximum bond dimension of $\chi = 6144$.

\subsubsection{Ground states of the Hubbard-Holstein model}
To simulate the Hubbard-Holstein model, we employ projected purification (PP)~\cite{Koehler2021}. In this approach, each physical bosonic site is augmented with a bosonic bath site. This construction allows for a global $U(1)$ particle-number conservation for all bosons, which would otherwise be broken by the interaction terms in the Hubbard-Holstein Hamiltonian. The resulting $U(1)$-symmetric block structure of the site tensors significantly simplifies MPS calculations~\cite{Singh2011}. We set the maximum number of phonons per physical and bath site to $d_{ph} = 48$, with the truncation error monitored via the eigenvalue spectrum of the one-body reduced density matrix (1-RDM) of the physical phonons, and construct the operators using finite-state machines~\cite{PAECKEL2017}. We use a bond dimension up to $\chi=6144$. 

\subsubsection{Dynamical spin structure factors and ARPES spectra}
We calculate the spectral function and dynamical structure factors from the time evolution of a state after a perturbation $\hat{c}_\mathrm{j\sigma}$ or $\hat{S}_\mathrm{j}$ respectively. Here, we describe the procedure for the spectral function. The dynamical structure factor calculations are done in analogy, replacing $\hat{c}_\mathrm{j\sigma}$ with $\hat{S}_\mathrm{j}$. 

The spectral function is defined as the Fourier transform of
\begin{align}
    A_{\vec{i}\vec{j}\sigma}(t) = \langle \psi \vert e^{i\mathcal{\hat{H}}t}\hat{c}^\dagger_{\vec{j}\sigma}e^{-i\mathcal{\hat{H}}t}\hat{c}_{\vec{i}\sigma}\vert \psi\rangle,
\end{align}
where $\psi$ is the ground state, $\hat{c}_{\vec{i}\sigma}^{(\dagger)}$ annihilate (create) particles at site $\vec{i}$ and $\sigma$ denotes the spin degree of freedom. In order to calculate this quantity, the ground state $\psi$ has to be calculated and $\hat{c}_{\vec{i}\sigma}$ is applied to create a hole. Then, the resulting state is time evolved using Krylov and the (two-site version of the) time dependent variational principle (TDVP) methods, see Ref.~\cite{PAECKEL2019167998}, and linear extrapolation~\cite{Barthel2009}. The spectral function is obtained by Fourier transformation w.r.t. spatial indices and time. For the latter, we use $A_\sigma(\vec{k},t)=A_\sigma(\vec{k},-t)^*$ (quasi-steady states) and hence
\begin{align}
    A_\sigma (\vec{k},\omega)= \frac{1}{2\pi}\int_0^{\infty} \mathrm{d}t\, \mathrm{Re}\left[ e^{i\omega t} A_\sigma(\vec{k},t)+ e^{-i\omega t} A_\sigma(\vec{k},-t)\right] = \frac{1}{\pi}\int_0^{\infty} \mathrm{d}t\, \mathrm{Re}\left[ e^{i\omega t} A_\sigma(\vec{k},t)\right].
\end{align}

We use a step size of $0.025t$ for an absolute time interval up to $20t$, with maximal bond dimension up to $\chi \leq 2048$. We use a small broadening of $\sigma_\omega=0.1t$ as in Ref.~\cite{Tang2024} and $\sigma_k=0.5\pi/L$ similar to Ref.~\cite{Tang2023}.

\FloatBarrier
\section{Additional results for ladder compounds}
\begin{figure*}[t]
\centering
\includegraphics[width=0.85\textwidth]{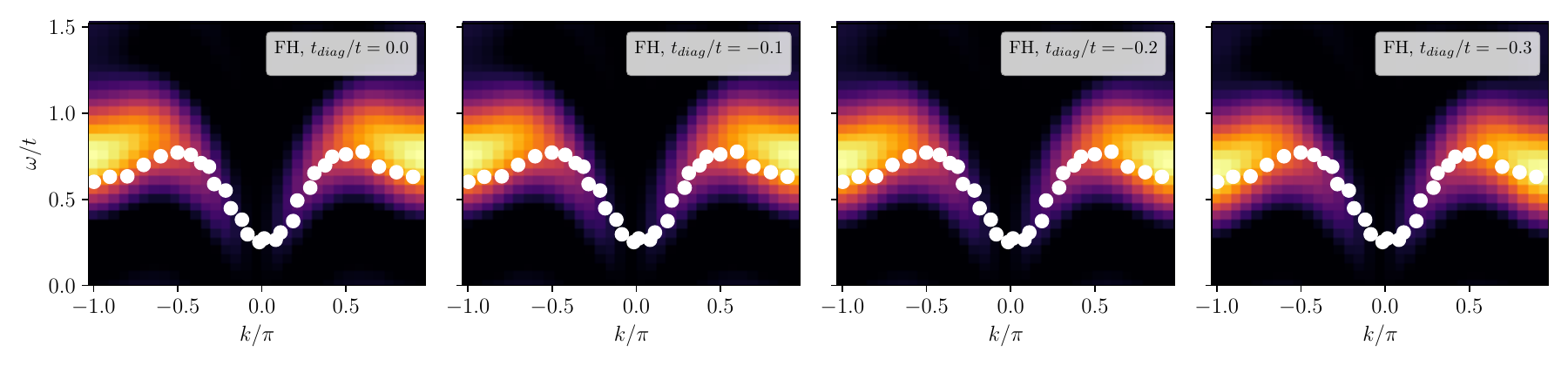}
\caption{Theoretical dynamical spin structure factor for ladders of length $L=33$ without doping for the Fermi Hubbard model with $V/t=0.0$ and different values of the diagonal hopping $t_{diag}$, compared to RIXS measurements from Ref. \cite{Padma2025} indicated by the white dots.}
\label{fig:tdiagscan0holes}
\end{figure*}

In the main text, we use the Fermi Hubbard parameters for the ladder compounds determined in Ref. \cite{Padma2025}, namely $U/t=(8-V)$, $t_\perp=0.84\;t$ and $t_{diag}/t=-0.3$. The latter value was comparison to the undoped case, where the downturn of the DSF at the zone boundary was found to be in best agreement with the experiment for $t_{diag}/t=-0.3$. This result is reproduced in Fig. \ref{fig:tdiagscan0holes}. Following Ref.~\cite{Padma2025}, this value was also used for the doped case in our work.\\

\begin{figure*}[t]
\centering
\includegraphics[width=0.95\textwidth]{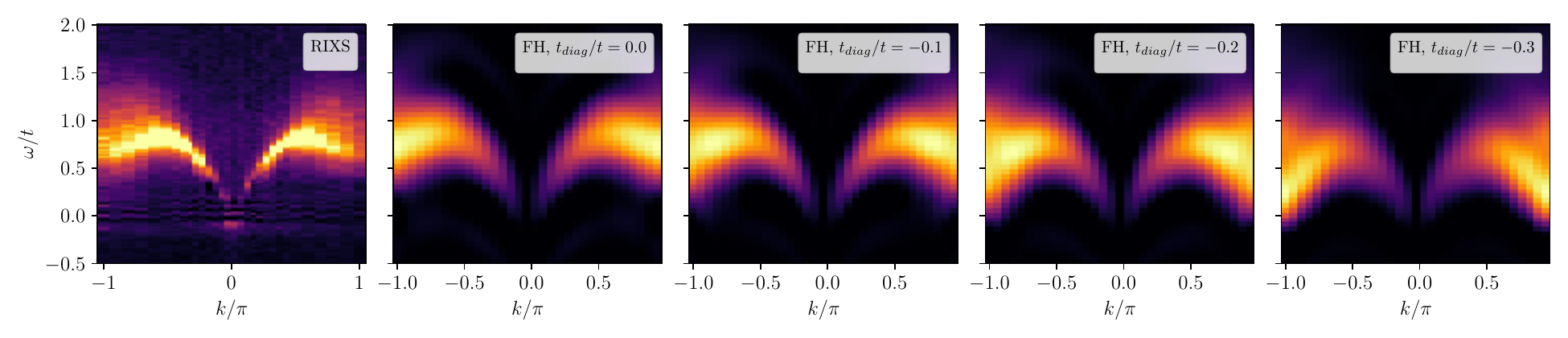}
\caption{Theoretical dynamical spin structure factor for ladders of length $L=33$ wit four holes for the Fermi Hubbard model with $V/t=0.0$ and different values of the diagonal hopping $t_{diag}$, compared to RIXS measurements from Ref. \cite{Padma2025} shown in the leftmost subfigure.}
\label{fig:tdiagscan4holes}
\end{figure*}

We emphasize, however, that for the doped case, other parameter combinations may also yield good agreement with experiment and could potentially require a smaller attractive interaction. This is suggested by the DSF for the doped case shown in Fig.~\ref{fig:tdiagscan4holes}: While the DSF for $t_{diag}/t=-0.3$ exhibits pronounced spin-flip weight, this weight is decreased for smaller values of $t_{diag}$. Further investigation is left to future work.

\section{Additional results for 1D compounds: ARPES Spectra}

\begin{figure*}[t]
\centering
\includegraphics[width=0.95\textwidth]{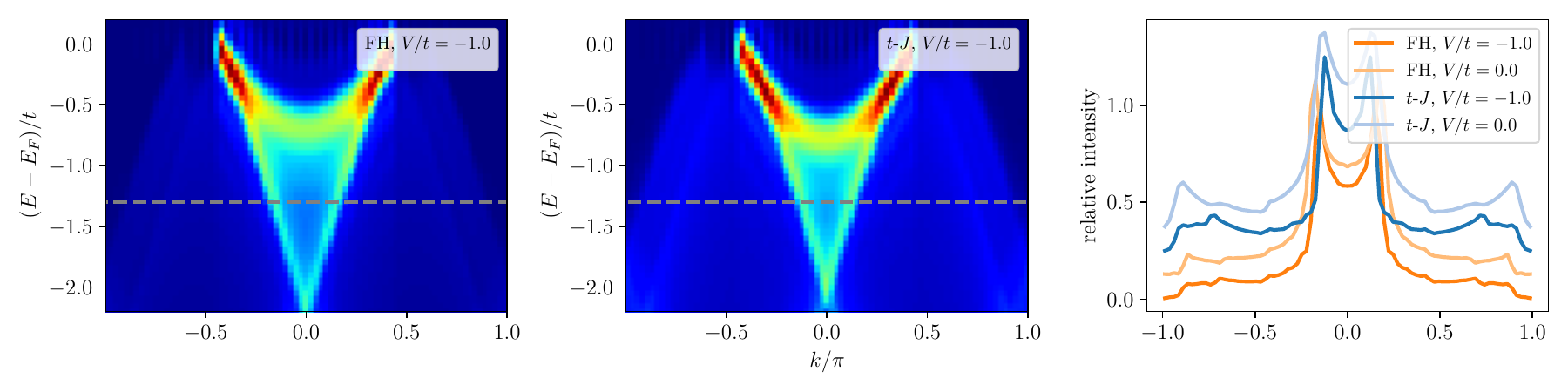}
\caption{Single-particle spectra for the Fermi-Hubbard model (left) and the $t$-$J$ model (middle), both with nearest neighbor attraction $V=-t$. For both models cuts at energies denoted by the gray lines are shown on the right and compared to the bare fermi Hubbard model (light orange). For all results $L=81$ and $\delta=10\%$.}
\label{fig:1DARPES}
\end{figure*}

Fig.~\ref{fig:1DARPES} shows the single-particle spectra for the Fermi-Hubbard and $t$-$J$ models with attraction, as well as a cut at energies indicated by the dotted lines. 

For both models, the spectra look very similar: Due to spin–charge separation in these 1D settings, spin and charge degrees of freedom decouple into independent excitations -- spinons and holons -- and propagate with distinct velocities and energy scales. The spinon dispersion is governed by the superexchange interaction $J=4t^2/U$, while the holon bandwidth is controlled by the hopping amplitude $t$. In the photoemission spectrum, these excitations give rise to separate spectral branches, which for a non-interacting system would only be present between $-k_F\leq k\leq k_F$: The spinon branch typically appears as an arc connecting $-k_F$ and $k_F$. The holon branch extends to lower energies.

In the case of interacting systems, the spectral weight extends beyond $-k_F\leq k\leq k_F$. This gives rise to the holon-folding (hf) branch emerging from $k=\pm k_F$. Additionally, another branch at $k=\pm3k_F$ appears. In general, one can observe that more spectral weight is distributed in the continuum for the $t$-$J$ model, and for the Fermi Hubbard models in the $3k_F$ branch. We have followed the discussion in Ref. \cite{Feiguin2023} here.\\

In the experiment \cite{Chen2021}, the intensity of the hf branch is higher than for the bare Fermi Hubbard model. In several works, it is shown that attractive interactions can enhance the intensity of this peak \cite{Chen2021,Tang2023,Feiguin2023,Tohyama2024}, see also our results in Fig.~\ref{fig:1DARPES} (right, dark orange vs. light orange). The $t$-$J$ model with attraction also features a pronounced hf branch, in agreement with Ref. \cite{Feiguin2023}. In general, the differences in intensity are much smaller than for the dynamical structure factor considered in the main text.

\end{document}